\documentclass[preprint,aps,showpacs,floatfix,superscriptaddress]{revtex4} 
\usepackage{graphicx}
\usepackage{subfigure}
\usepackage{array}
\usepackage{color}
\usepackage{amsmath}
\usepackage{amsxtra}
\usepackage{tablists}
\usepackage{diagbox}
\usepackage{amstext}
\usepackage{amssymb}
\usepackage{latexsym}
\usepackage{dsfont}
\usepackage{verbatim}
\usepackage{mathrsfs}
\usepackage{graphicx}
\usepackage{afterpage}
\usepackage[position=t,singlelinecheck=off]{subfig}
\usepackage{multirow}
\usepackage{marvosym}
\usepackage{floatrow}
\usepackage{textcomp}
\usepackage[utf8]{inputenc}
\newfloatcommand{capbtabbox}{table}[][\FBwidth]

\usepackage{cancel}
\usepackage{slashed}

\newcommand{\beq}{\begin{equation}\begin{split}}
\newcommand{\eeq}{\end{split}\end{equation}}
\newcommand{\bea}{\begin{eqnarray}}
\newcommand{\eea}{\end{eqnarray}}

\newcommand{\be}{\begin{equation}}
\newcommand{\ee}{\end{equation}}
\newcommand{\ba}{\begin{eqnarray}}
\newcommand{\ea}{\end{eqnarray}}
\def\bs{\begin{subequations}}
\def\es{\end{subequations}}
\def\a{\alpha}
\def\b{\beta}

\def\vr{\varrho}

\def\p{\partial}

\def\rmd{d}

\RequirePackage[normalem]{ulem}

\begin{document}
\title{Fractional quantum fields with Le'vy paths}
\author{Zheng-Wei Cheng}
\email{chengzw@snnu.edu.cn}
\affiliation{School of Physics $\&$ Information Technology, Shaanxi Normal University, Xi'an 710119, China}
\author{Xia Wan}
\email{wanxia@snnu.edu.cn}
\affiliation{School of Physics $\&$ Information Technology, Shaanxi Normal University, Xi'an 710119, China}
\author{You-Kai Wang}
\email{wangyk@snnu.edu.cn}
\affiliation{School of Physics $\&$ Information Technology, Shaanxi Normal University, Xi'an 710119, China}
\date{\today}

\begin{abstract}
We develop a path integral approach to quantum field theory that is defined over the paths of the Le'vy flights possessing a fractal dimension $1<d_f<2$. In standard quantum field theory, the fractality of the Brownian trajectories lead to a dispersion relation of quadric form. While the Le'vy paths lead fractional quantum field theory to a fractional dispersion relation. By considering Le'vy paths in time, we calculate density of states for a massless scalar field with box boundary condition. The density of states show behaviors dual to lower dimensional system, and the corresponding black body radiation has an energy spectrum dual to that in lower dimensional black body radiation. We derive the fractional equations of motion for scalar field, vector field and spinor field in zero temperature. Their propagators have been calculated. Based on above derivation, we calculate the one-loop self-energy of electron in Lev'y paths to show how this scheme works equally in renormalization as dimensional regularization does. Nevertheless, We found that gauge symmetry prevent Dirac spinor field from Le'vy paths and demonstrate that these paths in Dirac spinor field leds to unstable electrons, thus Le'vy paths are overcame by ordinary Brownian paths. While Le'vy paths are permitted in electromagnetic field by gauge symmetry. It led to a non-local phase and interaction with electron where electron charge is a composite quantity rather than a fundamental constant which maybe physically observable.
\end{abstract}
\maketitle
\section{INTRODUCTION}
Fractal was introduced to science firstly by Mandelbrot\cite{1982fgn..book.....M} to calculate the length of coastline. In physics, fractal was firstly used to describe the Brownian motion in which particle moving in nondifferentiable\cite{1982fgn..book.....M,Jens..book.....M}, self similar trajectories whose fractal dimension is different from topological dimension. In quantum physics, Feynman firstly used fractal to construct path integral approach of quantum mechanics. Feynman and Hibbs\cite{Feynmanbook} reconstruct nonrelativistic quantum mechanics as an integral over Brownian paths which is nonfractional. 

As Feynman path integral approach is also used in constructing quantum field theory, we give an extended fractality in quantum field theory through fractional path integral. Specifically, we construct a fractional path integral to develop fractional quantum field theory over paths of Le'vy flights.

The Brownian motion is a special case of Le'vy stochastic process which is based on Le'vy's theory of  stable probability distributions\cite{Gardiner1986HandbookOS,Levy1955ThorieDL}. The Le'vy distribution is stable to addition which thus can be seen as a generalization of Gaussian law. The Le'vy process can be characterized by Le'vy index $d, 0<d<2$, while $d=2$ case corresponds to Brownian motion. Le'vy process is widely used in various models: turbulence\cite{PhysRevA.35.3081}, chaotic dynamics\cite{10.1016/0167-2789(94)90254-2}, plasma physics\cite{10.1063/1.871453}, financial dynamics\cite{10.1038/376046a0}, biology and physiology\cite{WEST19941}, quantum mechanics\cite{Laskin_2000}, fractional Heisenberg equations\cite{Tarasov:2008}, quantum field theory and gravity for fractal spacetime\cite{Calcagni_2010,Calcagni_2012}, fractional Laplace and d'Alembert operators\cite{10.1007/BF02395016}, long-range memory and dissipative process\cite{Calcagni:2011kn}, multi-fractal spacetime\cite{Calcagni:2011kn,Calcagni:2021ipd}, lattice models with long-range properties\cite{Tarasov2006ContinuousLO,Tarasov2006MapOD,Tarasov2014TowardLF,Tarasov2015LatticeFC}.

We use fractional calculus to generalize fields equation from ordinary $d=2$ Brownian motion to $d=d_f$ Le'vy flights[see~Eq.(\ref{Le'vym})] , which is the origin of term "fractional quantum field theory".

The paper is organized as follows. In Sec.\ref{pil} we introduce fractional derivatives to Lagrangian of fields and illustrate how its fractal dimension vary with the order $\a$ of fractional derivatives and give a physical example on radiation of black body[see~Fig.(\ref{en})]. 

In Sec.\ref{Lagrangian}, we formulate fractional Klein-Gordon field, fractional electromagnetic field and fractional Dirac spinor field and derive the propagators and equations of motion. In Sec.\ref{calculation}, we calculate the electron self-energy and demonstrate it is consistent with Dimensional Regularization at $\a=1$ case. 

In Sec.\ref{gauge symmetry}, we demonstrate Le'vy paths in Dirac spinor field is unstable by studying its gauge symmetry and deduce to decay of electron. Besides, we find a nonlocal phase factor which corresponds to  nonlocal interaction[see Eq.~(\ref{ng}) and Eq.~(\ref{ni})] by studying gauge symmetry in fractional electromagnetic field and nonfractional Dirac spinor field. In the conclusion, we discuss some physical properties result from fractality of Le'vy paths.

\section{Path integral with Le'vy trajectories}
\label{pil}
We propose a path integral approach based on Le'vy flights to quantize field. It is realized by transformation ${\cal L}(\p_\mu\phi, \phi)\rightarrow{\cal L}(\p^\a_\mu\phi, \phi)$, where $\a$ denotes the order of derivative $\frac{d^\a}{dx^\a}$, $\mu$ denotes the direction of derivative $(t, x_1, x_2, x_3)$. We take scalar field in one dimension as an example to illustrate how it is related to Le'vy paths. Its path integral is 
\begin{equation}
    \int D_L\phi(x)\exp{[\int_{-\infty}^{+\infty}dx(-\frac{1}{2}m^2\phi^2)]}\,,
\end{equation}
where the Le'vy path integral measure $D_L\phi(x)$is 

\begin{align}
   D_L\phi(x)&=D\phi(x)\exp{[\int_{-\infty}^{+\infty}dx\frac{1}{2}(\p^\a\phi)^2]}\\&=D\phi(x)\exp{[\int_{-\infty}^{+\infty}\triangle x\frac{1}{2}(\frac{\triangle\phi}{\triangle x})^{2\a}]}\,.\label{Le'vym} 
\end{align}
It implies scale relation
\begin{equation}
    \triangle\phi\sim(\triangle x)^{\frac{2\a-1}{2\a}}\,.
\end{equation}
According to the definition of Hausdorff dimension, it posses a fractal dimension $d_f=\frac{2\a}{2\a-1}$. When $\a=1, d_f=2$, which is the typical dimension for Brownian path. While $\a\neq1$ is for Le'vy path. In declaration, when we use transformation $\mathcal{L}(\p_\mu\phi, \phi)\rightarrow{\cal L}(\p^\a_\mu, \phi)$, we substitute Brownian path by Le'vy path.

Next we consider a simple system, it is a massless scalar field in a box with length $L$ for every side. The Lagrangian is 
\begin{equation}
    {\cal L}=\frac{1}{2}\phi\p^\a_\mu\bar{\p}^\a_\mu\phi\,,\label{massless}
\end{equation}
where $\p^\a_\mu=(i\eta\p_t^\a, \p_1, \p_2, \p_3), \bar{\partial}^\alpha_\mu=(i\eta\bar{\partial}^{\alpha}_{t}, -\p_1, -\p_2, -\p_3)$, and for convenience of discussion, we restrict $\a<1$, $\eta$ is the generalized fractal
coefficient with dimension $m^{1-\a}$. And 
\begin{equation}
\begin{split}
   \partial^{\alpha}_tf(t)=\frac{1}{\Gamma(1-\alpha)}\int_{-\infty}^{t}\frac{dt'}{(t-t')^{\alpha}}\partial f(t')\label{liou}
\end{split}
\end{equation}
is Liouville fractional derivative, 
\begin{equation}
\begin{split}
    \bar{\partial}^{\alpha}_tf(t)=\frac{-1}{\Gamma(1-\alpha)}\int_{t }^{+\infty}\frac{dt'}{(t'-t)^{\alpha}}\partial f(t')
\end{split}
\end{equation}
is Weyl fractional derivative. They have relation 
\begin{equation}
    \int_{-\infty}^{\infty}dxf\partial^{\alpha}g
=\int_{-\infty}^{\infty}dxg\bar{\partial}^{\alpha}f\,.\label{partin}
\end{equation}
One reason we choose them is that they have good property 
\begin{equation} 
    \partial^{\alpha}e^{\lambda x}=\lambda^{\alpha} e^{\lambda x}\,,\qquad
\bar{\partial}^{\alpha}e^{\lambda x}=(-\lambda)^{\alpha} e^{\lambda x}\,.
\end{equation}
The fractional derivative is a convolution of field strength $\p\phi$ and a time scale function 
\begin{equation}
 \frac{\eta(\triangle t)^{-\a}}{\Gamma(1-\alpha)}\,.
 \end{equation}
It can be interpreted as a induced field strength. To make it clear, we can rewrite Eq.~(\ref{liou}) as
\begin{align}
    \eta\partial^{\alpha}_tf(t)&=\frac{\eta}{\Gamma(1-\alpha)}\int_{-\infty}^{t}\frac{dt'}{(t-t')^{\alpha}}\partial f(t')\\&=\p f(t)+\int_{-\infty}^{t}dt'\p f(t')[\frac{\eta}{\Gamma(1-\alpha)(t-t')^\a}-\delta(t-t')]\,,
\end{align}
The second term is a time average with distribution $\frac{\eta}{\Gamma(1-\alpha)(t-t')^\a}-\delta(t-t')$ over whole past time. If we transform to frequency space
\begin{align}
    \eta\partial^{\alpha}_tf(t)&=\int d\omega \eta(i\omega)^{\a-1}f(\omega)\p\exp{(i\omega t)}\\&=\int d\omega f'(\omega)\p\exp{(i\omega t)}\,,
\end{align}
where $f'(\omega)=\eta(i\omega)^{\a-1}f(\omega)$, the time average will give a modification to spectra function $f(\omega)$ which will reflect to dispersion relation and ultimately induce noticeable changes to the density of state.

It is easy to obtain the propagator
\begin{equation}
  C\sum_{n1,n2,n3}\int d\omega\frac{-ie^{-ik\cdot(x_a-x_b)+i\omega t}}{\eta^2\omega^{2\a}-k_{n1}^2-k_{n2}^2-k_{n3}^2}=G(x_a-x_b)  
\end{equation}
from Eq.~(\ref{massless}), $C$ is a constant.

\begin{figure}[H]
	\centering
		\includegraphics[width=8cm, height=6cm]{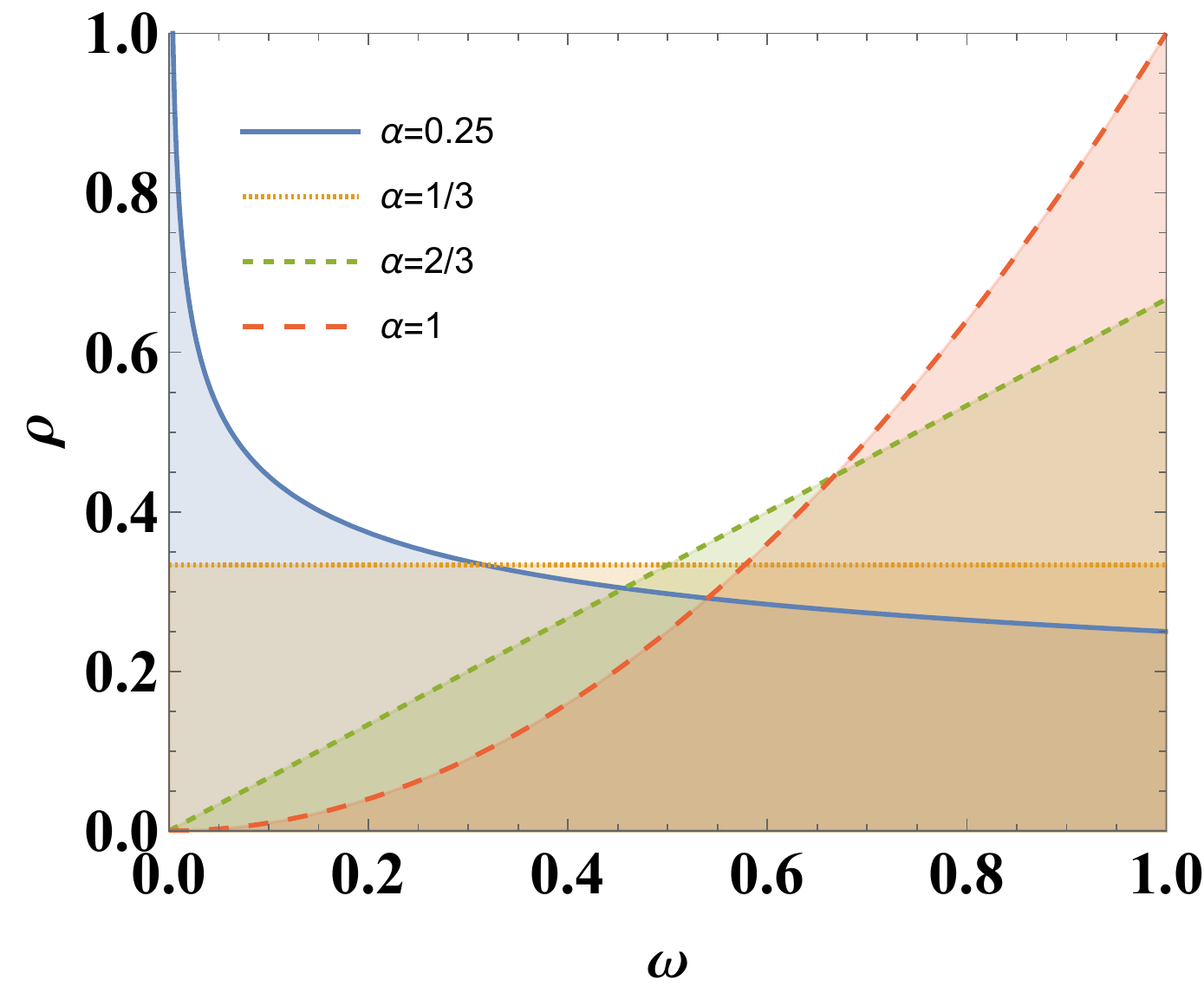}
		\caption{\centering\textcolor{black}{Dos $\rho(\omega)$ for different value of $\a$.}}
  \label{dos}
\end{figure}

We can obtain its momentum $(\eta\omega^\a, \frac{n1}{L}, \frac{n2}{L}, \frac{n3}{L})$ and dispersion relation $\eta^2\omega^{2\a}=k_n^2$. The phase velocity $v=\frac{\omega}{k}=\eta^{-1}\omega^{1-\a}$ and group velocity $v_p=\frac{d\omega}{dk}=\a\eta^{-2}\omega^{1-2\a}$ both depend on frequency. We can also obtain the density of states(dos) from dispersion relation and boundary condition.
\begin{equation}
\rho(\omega)=c\a\omega^{3\a-1}=\frac{4\a}{3\eta^3}L^3\pi \omega^{3\a-1}\,.
\end{equation}
There are three special cases for $\a=\frac{1}{3}, \frac{2}{3}, 1$. we plot them in Fig.~\ref{dos} for $C=1$.
When $\a=1$, the dos $\rho(\omega)$ is just that of ordinary three dimension scalar field. When $\a=\frac{2}{3}$, $\rho(\omega)$ behave as a two dimension system. When $\a=\frac{1}{3}$, it convergent to one dimension system. More interesting things happen when $\a<\frac{1}{3}$ that $\rho(\omega)$ shows a property between zero and one dimension system with a noticeable singularity at $\omega=0$ which means a divergent degeneracy. 

\begin{figure}[H]
	\centering
		\includegraphics[width=8cm, height=6cm]{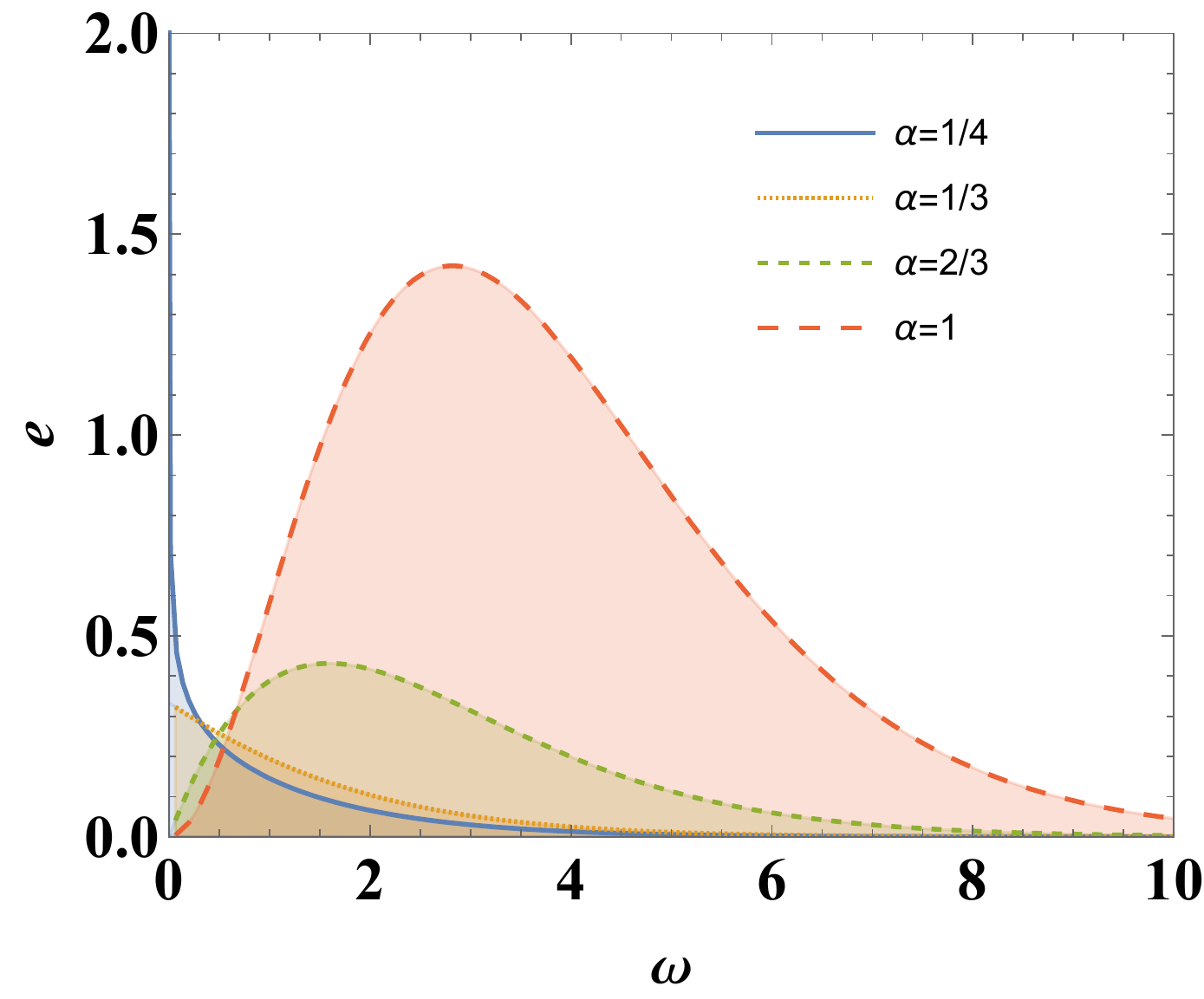}
		\caption{\centering\textcolor{black}{Energy spectrum $e(\omega)$ of black body radiation.}}
  \label{en}
\end{figure}

As an example, we plot the corresponding energy spectrum $e(\omega)$ of black body radiation in Fig.~\ref{en}.
When $\a=1$, the energy spectrum $e(\omega)$ is just that of radiation from three dimensional black body. When $\a=\frac{2}{3}$, $e(\omega)$ behave as a two dimensional black body. When $\a=\frac{1}{3}$, it convergent to one dimensional black body. While when $\a<\frac{1}{3}$, $e(\omega)$ shows a property between zero and one dimensional system with a divergent energy density e at $\omega=0$, but this divergence will not led divergent radiation power. 

\section{Same approach to quantum field in zero temperature}
\label{Lagrangian}
For simplification, we consider Klein-Gordon field, electromagnetic field and Dirac spinor field.
%In this section, firstly we introduce fractional derivative in the Lagrangian for scalar, vector and spinor fields and get their equations of motion, then we derive the fractal propagators.
To treat space and time in as a same manner as possible, we choose $\p^\a_\mu=(\p^\a_0, \p^\a_1, \p^\a_2, \p^\a_3)$, $\bar{\p}^\a_\mu$ has the similar form.

As we know in ordinary Dirac equation, the kinetic operator is $i\slashed{\p}-m$, where $\slashed{\p}=\gamma^\mu\p_\mu$. In our approach, ${\cal L}(\p_\mu\phi, \phi)\rightarrow{\cal L}(\p^\a_\mu\phi, \phi)$. So the Lagrangian for Dirac spinor field is 
\begin{equation}
{\cal L}=\frac{1}{2}\bar{\psi}(i\eta\slashed{\p}^\a-m)\psi\,,\label{dirac}
\end{equation}
where $\eta$ is a generalized fractal
coefficient with dimension $m^{1-\a}$. 

We can derive the equation of motion by
\begin{equation}
    \begin{split}
        0&=\delta S_1\\&=\int d^4x[\frac{\p {\cal L}}{\p\phi}\delta\phi+\frac{\p {\cal L}}{\p(\p_\mu^\a\phi)}\delta(\p_\mu^\a\phi)]\\&=\int d^4x[\frac{\p {\cal L}}{\p\phi}\delta\phi+\bar{\p}_\mu^\a(\frac{\p {\cal L}}{\p(\p_\mu^\a\phi)})\delta\phi]\,,
    \end{split}
\end{equation}
where in last step we use Eq.~(\ref{partin}). The Euler-Lagrange equation
\begin{equation}
    \frac{\p {\cal L}}{\p\phi}+\bar{\p}_\mu^\a(\frac{\p {\cal L}}{\p(\p_\mu^\a\phi)})=0\,.
\end{equation}
We obtain the equation of motion for Dirac field
\begin{equation}
    (i\eta\slashed{\p}^\a-m)\psi=0\,,\qquad(i\eta\bar{\slashed{\p}}^\a-m)\bar{\psi}=0\,,
\end{equation}
They describe positive time direction and negative time direction electron respectively. By adopting similar path integral approach in Section~(\ref{pil}), we obtain the propagator
\begin{equation}
\begin{split}
	\int\frac{d^4k}{(2\pi)^4}\frac{i\eta (-i\slashed{k})^\a+m}{-\eta^2(-ik)^{2\alpha}-m^2+i\epsilon}e^{-ik\cdot(x_1-x_2)}\,.
\end{split}
\label{spinorpropagator}
\end{equation}

Since in $\a=1$, Dirac spinor field and Klein-Gordon field have relation 
\begin{equation}
  -(i\slashed{\p}\phi+m\phi)(i\slashed{\p}\phi-m\phi)=(\p_\mu\phi)^2+m^2\phi^2\,,\label{delanm}  
\end{equation}
where $\gamma^\mu\p_\mu\gamma^\nu\p_\nu=\p^2$, naturally we propose  
\begin{equation}
    -(i\eta\slashed{\p}^\a\phi-m\phi)(i\eta\slashed{\p}^\a\phi+m\phi)=(\eta\p^{\a}_\mu\phi)^2+m^2\phi^2\,.\label{klein}
\end{equation}
One can easily verify that when $\a\rightarrow1$, it recovers to Eq.~(\ref{delanm}).

According to  Eq.~(\ref{klein}), the Lagrangian for Klein-Gordon field is:
 \begin{equation}
 \begin{split}
  {\cal L}=\frac{1}{2}[(\eta\p^\a_\mu\phi)^2+m^2\phi^2]\,. \label{fc}  
 \end{split}
 \end{equation}
Here $d_{f}=\frac{2\alpha}{2\alpha-1}$. Consider the case $d_{f}>0$ (or equivalently $\a>\frac{1}{2})$) for the requirement of the first moment’s existence, therefore we restrict $\alpha\in(\frac{1}{2},1]$.

In same manner, its equation of motion is
\begin{equation}
 \eta^2\p^{\a}\bar{\p}^\a\phi+m^2\phi=0\,. 
\end{equation}
By similar path integral approach in Section~(\ref{pil}), we obtain the propagator
\begin{equation}
\begin{split}
	\left\langle 0\left| T\phi(x_1)\phi(x_2)\right|0\right\rangle=\int\frac{d^4k}{(2\pi)^4}\frac{-ie^{-ik\cdot(x_1-x_2)}}{\eta^{2}k^{2\alpha}+m^{2}+i\epsilon}=D_F(x_1-x_2)\,.
\end{split}
\label{propagator}
\end{equation}
Finally, we write down the Lagrangian ${\cal L}(\p^\a_\mu\phi, \phi)$ for electromagnetic field
\begin{equation}
\begin{split}
	{\cal L}=-\frac{1}{2}\eta^{2}A^\mu(\partial^{\alpha}_{\lambda}\bar{\partial}^{\alpha}_{\lambda} g_{\mu\nu}-\partial^\alpha_\mu\bar{\partial^\alpha_\nu})A^\nu\,.\label{fv}
\end{split}
\end{equation}
Its equation of motion is 
\begin{equation}
    \bar{\p}^\a\p^\a A_\nu-\p_\nu^\a\bar{\p}^{\mu\a}A_\mu=0\,.\label{eomv}
\end{equation}
One can easily verify that it is invariant under a non-local gauge transformation
\begin{equation}\label{31}
 A_\mu(x)\rightarrow A_\mu(x)-\frac{\eta}{e}\partial_\mu^\a\a(x)\,,
\end{equation}
which we will further discuss it in later section. The propagator with Feynman gauge is
\begin{equation}
\begin{split}
	\frac{ig_{\mu\nu}}{\eta^{2}k^{2\alpha}+i\epsilon}\,.
\end{split}
\end{equation}
Based on above approachs, only propagators are changed in case of Le'vy paths. The interacting is unaffected. Therefore, vertexes in path integral are the same as the ordinary cases. In next section we will illustrate that it will led to a result similar to the Dimensional Regularization in one-loop diagram. 
\section{The electron Self-Energy}
\label{calculation}
Here we take the one-loop electron self-energy process as an example to demonstrate how to calculate the ultraviolet divergence in this approach.

In this one-loop process we need to do such integral
\begin{equation}
\begin{split}
	\label{eq2}I=\int_{-\infty}^{+\infty} d^4k\frac{-2i\eta(-i\slashed{k})^\a+4m}{-\eta^2(-ik)^{2\a}-m^2+i\epsilon}\frac{1}{\eta^2(k-p)^{2\a}+i\epsilon}\,,
\end{split}
\end{equation}
Its ordinary form when $\a\rightarrow1$ is commonly solved by using Feynman parameterization method 
\begin{equation}
\begin{split}
	&\frac{1}{k^2+m^2+i\epsilon}\frac{1}{(k-p)^2+i\epsilon}\\=&\int_{0}^{1}dx\frac{1}{[(1-x)(k^2+m^2+i\epsilon)+x((k-p)^2+m^2+i\epsilon)]^{2}}\\=&\int_{0}^{1}dx\frac{1}{[l^2+x(1-x)p^2+(1-x)m^2+i\epsilon]^2}\,,
\end{split}
\end{equation}
where $l=(k-xp)^2$. In ordinary case $\a=1$, it is easy to do such transformation as $(k-p)^2=k^2+p^2-2kp$. While in Le'vy path, $(k-p)^2\rightarrow(k-p)^{2\a}$, this decomposition fail. To address this problem, we can use formulae\cite{Calcagni:2011kn,https://doi.org/10.1118/1.594867}
\begin{equation}
\begin{split}
 \partial^{\alpha}(x-c)^{\beta}
&=(-1)^{\alpha}\bar{\partial}^{\alpha}(x-c)^{\beta}\\
&=(-1)^{\alpha}\frac{\Gamma(\beta+1)}{\Gamma(\beta-\alpha+1)}\frac{\sin(\pi\beta)}{\sin[\pi(\beta-\alpha)]}(x-c)^{\beta-\alpha}\,,
\end{split}
\label{eq1}
\end{equation}
where $c$ is a constant. Using this equation and the property that $\p^\a, \bar{\p}^a$ are linear operators, we can derive following formulae
\begin{equation}
\begin{split}
	\bar{\partial}^{2-2\alpha}(k-p)^2&=\frac{2}{\Gamma(2\alpha+1)}\frac{\sin(2\pi)}{\sin(2\alpha\pi)}(k-p)^{2\alpha}\\&=\bar{\partial}^{2-2\alpha}k^2-\bar{\partial}^{2-\alpha}2kp+\bar{\partial}^{2-2\alpha}p^2\\&=\frac{2}{\Gamma(2\alpha+1)}\frac{\sin(2\pi)}{\sin(2\alpha\pi)}k^{2\alpha}-\frac{1}{\Gamma(2\alpha)}\frac{\sin(\pi)}{\sin[(2\alpha-1)\pi]}2pk^{2\alpha-1}+b,
    \\ &\mbox{where} \ \ \  b=\left\{\begin{array}{c}
		0\quad \alpha\neq1\\p^{2}\quad\alpha=1.
	\end{array}
 \right.	
\end{split}
\end{equation}
Out of expectation, it has a rather simple form after reduction
\begin{equation}
\begin{split}
	(k-p)^{2\alpha}=k^{2\alpha}-2\alpha pk^{2\alpha-1}+b,\quad b=\begin{cases}0\quad&\alpha\neq1\\p^{2}\quad&\alpha=1,
		\end{cases}
\end{split}
\end{equation}
and 
\begin{equation}
\begin{split}
	(k-p)^{\alpha}=k^{\alpha}-c,\quad c=\begin{cases}0\quad&\alpha\neq1\\p\quad&\alpha=1.
		\end{cases}
\end{split}
\end{equation}
The calculation can be seen in Appendix~\ref{ioop}. We just present result here
\begin{equation}
\begin{split}
	I=2(-i)^{-\a}\eta^{-1}(i\slashed{z}^{-1}I_3-2m_1I_2)\,,
\end{split}
\end{equation}
where $m_1=m/(\eta (-i)^\a)$ and $\slashed{z}=\gamma^{\mu}n_{\mu}$, $n_{\mu}=z$ is a unit vector along the $z$ axis. And
\begin{equation}
\begin{split}
	I_2=\frac{e^{\frac{i\pi}{2\alpha}}}{2\alpha}\int d\Omega(\xi^{-\frac{2}{\alpha}})\triangle^{\frac{2-2\alpha}{\alpha}}\frac{\Gamma(\frac{2\alpha-2}{\alpha})\Gamma(\frac{2}{\alpha})}{\Gamma(2)}\,,
\end{split}
\end{equation}
\begin{equation}
\begin{split}
	I_3=\frac{1}{2\alpha}\int d\Omega(\cos^{\alpha}\theta \xi^{-\frac{4+\alpha}{2\alpha}})\triangle^{\frac{4-3\alpha}{2\alpha}}\frac{\Gamma(\frac{3}{2}-\frac{2}{\alpha})\Gamma(\frac{1}{2}+\frac{2}{\alpha})}{\Gamma(2)}+b_3I_2\,,
\end{split}
\end{equation}
where $\triangle=x(1-x)c+(1-x)m_1^2, b_3=xp_3$, and 
$$\xi=(\cos\theta)^{2\alpha}+(\sin\theta)^{2\alpha}[(\cos\phi)^{2\alpha}+(\sin\phi)^{2\alpha}((\cos\omega)^{2\alpha}+(\sin\omega)^{2\alpha})].$$
Let $\a\rightarrow1, I_3=xp_3I_2$, this integral is reduced to 
\begin{equation}
    \begin{split}
        I=2i\pi^2[2m-x\slashed{p}]\Gamma(0)\,,
    \end{split}
\end{equation}
which is exactly the same as the result of Dimensional Regularization approach.

\subsection{$\alpha=1$ case and the Passarino-Veltman Reduction}
From the previous calculation we can see that the angle integral is not easy unless the dynamic related dimensional parameter $\alpha=1$. In this special case we get the result consistent with that from the Dimensional Regularization. So in this subsection we will discuss if it can work as an substitution of Dimension Regularization in the general loop integral.

In four dimension spacetime, according to Passarino-Veltman Reduction, an one-loop integral can be expressed as a linear combination of multi-point scalar functions. Therefore, to obtain an equivalent result to that of the dimensional regularization, we only need to calculate the one to four points scalar functions in the framework of our fractal path integral with $\a=1$.  

For an one-loop integral with $N$ propagators, the number of its outlegs is also $N$. The general one-loop integral without Dimensional Regularization can be wrote in a unitary form~\cite{Denner:1991kt} 
\begin{equation}
\begin{split}
	T^{N}_{\mu_{1}\cdots\mu_{p}}(p_{1},\cdots,p_{N-1},m_{0},\cdots,m_{N-1})= \frac{1}{i\pi^{2}}\int d^{4}q\frac{q_{\mu_{1}}\cdots q_{\mu_{p}}}{D_{0}D_{1}\cdots D_{N-1}},
\end{split}
\end{equation}
where $D_{0}=q^{2}-m_{0}^{2}+i\epsilon$,   $D_{i}=(q+p_{i})^{2}-m_{i}^{2}+i\epsilon~(i=1,\cdots, N-1)$, $\mu$ is Lorentz index, $m_i~(i=0,1,\cdots, N-1)$ is mass of the $i$-th particle in the propagator, $q$ is the loop momentum. $q_{\mu_{ i}}~(i=1,\cdots,p)$ is the momentum in the numerator. As the numerator of the propagator for massless photon is 1, the number $p$ of momentum $q_{\mu_{i}}$ in the numerator should be equal or less than $N$. The momentum of the $i$th outleg is $p_{i}-p_{i-1}~(i=1,\cdots,N-1,p_0=0)$.

After the process of Passarino-Veltman Reduction~\cite{Denner:1991kt}, $T^{N}$ with $N>4$ can be expanded as a linear combination of $T^{1},T^{2},T^{3}$, and $T^{4}$. To calculate them parallelly in the scheme of fractal path integral, just substitute the original  momentum by the fractal momentum and repeat the process we do in the example calculation of the electron self-energy. Same result can be obtained as that by Dimension Regularization.

\section{The gauge symmetry}\label{gauge symmetry}
In last section, we use the Le'vy path integral' property in $\a\rightarrow1$ to specify the divergence in standard model quantum field. While in this section, we will discuss the $\a\neq1$ case from perspective of gauge symmetry to illustrate that Le'vy paths are only physically permitted in electromagnetic field. 
\subsection{Le'vy paths in Dirac spinor field are unstable and are forbidden by Gauge symmetry}
In ordinary Brownian path integral $\a=1$, the Lagrangian of quantum electrodynamics 
\begin{equation}
  {\cal L}=\bar{\psi}(i\slashed D-m)\psi-\frac{1}{4}(F_{\mu\nu})^2
\end{equation}
where $D_\mu=\p_\mu+ieA_\mu(x)$ is invariant under gauge transformation $A_\mu(x)\rightarrow A_\mu(x)-\frac{1}{e}\a(x), \psi(x)\rightarrow e^{i\a(x)}\psi(x)$. In Le'vy path integral, however, the Dirac spinor field Lagrangian Eq.~(\ref{dirac}) fail to give right counter-term 
\begin{equation}
  \slashed{\p}^\a (e^{i\a(x)}\psi(x))\rightarrow e^{i\a(x)}\slashed{\p}^\a\psi(x)+ie^{i\a(x)}\slashed{\p}^\a\a(x)\psi(x)  
\end{equation}
because of property\cite{Calcagni:2011kn} 
\be\label{leru}
\p^\a(fg)=\sum_{j=0}^{+\infty}\binom{\a}{j} (\p^j f)(\p^{\a-j}g)\,,\qquad \binom{\a}{j}=\frac{\Gamma(1+\a)}{\Gamma(\a-j+1)\Gamma(j+1)}\,.
\ee
We can try to calculate it in definition Eq.~(\ref{liou})
\begin{align}
    \p^{\a}(e^{i\a(x)}\psi(x))&=\int_{-\infty}^{x} \rmd\vr_{1-\a}(x')\,\p(e^{i\a(x')}\psi(x'))\\&=\int_{-\infty}^{x} \rmd\vr_{1-\a}(x')\,i\p\a(x')e^{i\a(x')}\psi(x')+\int_{-\infty}^{x} \rmd\vr_{1-\a}(x')\,e^{i\a(x')}\p\psi(x')\,,\label{44}
\end{align}
where $\rmd\vr_{1-\a}(x')=\rmd x'\frac{(x-x')^{-\a}}{\Gamma(1-\a)}$. Using $y=x-x'$, the integral become
\begin{align}
   \partial^{\alpha}(e^{i\a(x)}\psi(x))= \ &\frac{1}{\Gamma(1-\a)}\int_{0}^{+\infty} \rmd y y^{-\a} \p_{x-y}(e^{i\a(x-y)}\psi(x-y))\\ \stackrel{\epsilon\rightarrow 0}{=}&\frac{1}{\Gamma(1-\a)}\int_{0}^{+\infty} \rmd y y^{-\a} (y-\epsilon)^{-1}y\p_{x-y}(e^{i\a(x-y)}\psi(x-y))\\ \stackrel{\epsilon\rightarrow 0}{=}&\frac{1}{\Gamma(1-\a)}\int_{0}^{+\infty} \rmd y y^{-\a} (y-\epsilon)^{-1}[\p_{x-y}(ye^{i\a(x-y)}\psi(x-y))\\ &+e^{i\a(x-y)}\psi(x-y)].          \label{gradient} 
\end{align}
Then use the formulae\cite{Complex}
\begin{align}
    \int^{+\infty}_{0}x^{\a-1}f(x)dx=\frac{2\pi i}{1-e^{2\pi\a i}}\sum R[z^{\a-1}f(z),z_j],\quad 0<\a<1\,,
\end{align}
where $R$ means residue.
If we assume $y\p_{x-y}(e^{i\a(x-y)}\psi(x-y))$ is analytic over the entire complex plane, we obtain
\begin{align}
    \partial^{\alpha}(e^{i\a(x)}\psi(x))\stackrel{\epsilon\rightarrow 0}{=}\frac{2\pi i\epsilon^{1-\a}}{e^{2\pi(1-\a) i}-1}\p_{\epsilon+x}(e^{i\a(\epsilon+x)}\psi(\epsilon+x))\,.
\end{align}
If $\a=1$, $e^{2\pi(1-\a) i}-1=0$, it will go back to $-\partial(e^{i\a(x)}\psi(x))$. However in Le'vy path $\a\neq1$, the $\epsilon$ will inevitably make $e^{i\a(x)}\psi(x)$ vanish unless $e^{i\a(x)}\psi(x)$ discontinuous at every $x$. Therefore, destroying of gauge symmetry by Eq.~(\ref{leru}) is inevitable. In other words, the gauge symmetry forbid all $\a\neq1$ Le'vy paths in Dirac spinor field. 

Further more, if we consider the Lorenz transformation of $\p^\a_\mu\stackrel{\Lambda}{\rightarrow}{\p'}^\a_\mu$
\begin{align}
    {\p'}^\a_\mu&=\frac{1}{\Gamma(1-\a)}\int_{-\infty}^{x^\nu}\Lambda^\mu_\rho dx^{'\rho}(\Lambda^\nu_\mu)^{1+\a}[(x-x')^{-\a}\p]_\nu\\&=(\Lambda^{\nu}_\mu)^{\a}\frac{1}{\Gamma(1-\a)}\int_{-\infty}^{x^\nu}dx^{'\nu}(x^\nu-x^{'\nu})^{-\a}\p_\nu\\&=(\Lambda^{\nu}_\mu)^{\a}\p^\a_\nu\,,
\end{align}
here we use $\int_{-\infty}^{x^\mu}dx^{'\mu} f(x')=g^\mu(x)$. Since $g^\mu(x)\stackrel{\Lambda}{\rightarrow}\Lambda^\nu_\mu g^\mu(x)$ and $f(x')$ is a scalar field. $\int_{-\infty}^{x^\mu}dx^{'\mu}\stackrel{\Lambda}{\rightarrow}\int_{-\infty}^{x^\nu}\Lambda^\nu_\mu dx^{'\mu}$. For Klein-Gordon field and electromagnetic field, the kinetic term
\begin{align}
    \p^\a_\nu\p^{\a\nu}\stackrel{\Lambda}{\rightarrow}(\Lambda^{\mu}_\nu)^{\a}(\Lambda^{\nu}_\lambda)^{\a}\p^\a_\mu\p^{\a\lambda}=(\delta^\mu_\lambda)^\a\p^\a_\mu\p^{\a\lambda}\,,
\end{align}
is Lorenz invariant. While for Dirac spinor field, that is $\gamma^\nu\p^\a_\nu\psi$
\begin{align}
  \gamma^\nu\p^\a_\nu\psi\stackrel{\Lambda}{\rightarrow}\gamma^\nu(\Lambda^{\mu}_\nu)^{\a}\p^\a_\mu D(\Lambda)\psi&=D(\Lambda)D^{-1}(\Lambda)\gamma^\nu D(\Lambda)(\Lambda^{\mu}_\nu)^{\a}\p^\a_\mu\psi\\&=D(\Lambda)[\Lambda^{\nu}_\lambda\gamma^\lambda(\Lambda^{\mu}_\nu)^{\a}\p^\a_\mu\psi]\,,
\end{align}
where $D(\Lambda)$ is the spinor representation of Lorenz transformation. It is not Lorenz invariant except $\a=1$, either if one can find a $\psi$ transforming as $\psi\stackrel{\Lambda}{\rightarrow}D(\Lambda^\a)\psi$. Besides, fractional Dirac spinor particle is unstable. Considering its propagator $\frac{i\eta (i\slashed{k})^\a+m}{-\eta^2(ik)^{2\alpha}-m^2}$, the scatting amplitude gets contribution from propagator: $S\sim\frac{1}{(ik)^{2\alpha}+\mu^2}$. To see its property in time, we transform it into time space and center-of-mass frame
\begin{align}
    \int dk_0\frac{i\eta \gamma^0(ik)_0^\a+m}{(i\eta(ik_0)^\a-m)(i\eta(ik_0)^\a+m)}e^{-ik_0t}\,.
\end{align}
%the interaction between fractal spinor field and fractal vector field is not gauge invariant because gauge transformation makes the gradient term vanish. After these calculations we fail to keep gauge symmetry for fractal spinor field.
It is obvious that the amplitude has two complex pole $k_1, k_2$ satisfying $i\eta(ik_1)^\a-m=0, i\eta(ik_2)^\a+m=0$, which means particle and antiparticle. Since $\a$ is a non-integer, the complex pole is inevitable, which means we can not make it real by choosing suitable $\eta$. They have contribution $S\sim e^{-ik_1t}, S\sim e^{-ik_2t}$. The decay rate is 
\begin{equation}
    |S|^2\sim e^{2Im(k_1)t}\,.
\end{equation}
Dirac spinor field with Le'vy paths will led to exponential decay of electron, which is nonphysical. Thus these paths are overcame by ordinary Brownian paths. The electron decay here results from structure of propagator which has complex poles suggesting invalidation of particle method. The complex poles here originate from non-integer order of $k_0$: $\a$, so when $\a\in N$ the poles will be real, hence no decay. Since $d_f=\frac{2\a}{2\a-1}$, it is obvious that only when $\a=1$ we have $d_f\in N, \a\in N$, which means Brownian path is only one to led both integer fractal dimension and stable particle. Besides, there is a interesting case when $\a=\frac{1}{2}$, in which the fractal dimension $d_f=\infty$ and the two pole degenerate to one, suggesting a massive particle with no antiparticle.     

\subsection{Non-local gauge phase and interaction in Le'vy paths electromagnetic field}
From above discussion, it is showed that fractional spinor field with L'evy paths is forbidden by gauge symmetry. In this part, we will demonstrate Le'vy paths in electromagnetic field are permitted. 

We introduce a non-local phase
\begin{align}\label{ng}
  e^{ic_1\gamma^\nu(\gamma I^{1-\b})_\nu\a(x)}\,,  
\end{align}
and a non-local electromagnetic interaction
\begin{align}\label{ni}
  iec_2\bar{\psi}(x)\gamma^\mu\psi(x)\gamma^\nu(\gamma\p^{\b-\a})_\nu A_\mu(x)\,.
\end{align} 
where $\gamma^\nu$ is $(\gamma^0,\gamma^1,\gamma^2,\gamma^3)$, $(\gamma I^{1-\b})_\nu$ is $(\gamma^0I^{1-\b}_0,\gamma^1I^{1-\b}_1,\gamma^2I^{1-\b}_2,\gamma^3I^{1-\b}_3)$, and $I^{1-\b}_\mu$ means a fractional integral over $x_\mu$, $c_1$ has dimension $m^{1-\b}$. $c_2$ has dimension $m^{\a-\b}$, $c_1, c_2$ are two generalized fractal factors and they have relation $c_1=c_2\eta$. It has a modification to electron charge $e=c_2e_0$, where $e_0$ is the bare unit electric charge, which means electron charge is a composite quantity. There is a special case when $\a=\b$, where $\frac{e}{e_0}=c_2$ is a pure number and interaction become local. The Le'vy paths electromagnetic field can have a gauge interaction as following,
\begin{align} 
    \bar{\psi}\slashed\partial\psi(x)&\rightarrow \bar{\psi}\slashed\partial\psi(x)+ic_1\slashed\partial^\b\a(x)\bar{\psi}\psi(x)\,,\\
    iec_2\bar{\psi}(x)\gamma^\mu\psi(x)\gamma^\nu(\gamma\p^{\b-\a})_\nu A_\mu(x)&\rightarrow iec_2\bar{\psi}(x)\gamma^\mu\psi(x)\gamma^\nu(\gamma\p^{\b-\a})_\nu A_\mu(x)-ic_1\slashed\p^\b\a(x)\bar{\psi}\psi(x)\,,
\label{CHint}
\end{align}
where we use the formulae\cite{Calcagni:2011kn}
\ba
\p^\a\,I^\b &=& I^\b\,\p^\a=\p^{\a-\b}\,,\qquad \forall~ \a,\b>0\,.
\ea
and $\gamma^\mu\gamma^\nu\p_\mu(\gamma I^{1-\b})_\nu=g^{\mu\nu}\p_\mu(\gamma I^{1-\b})_\nu=\gamma^\mu\p^\b_\mu=\slashed\p^\b$, which can be derived through the commutation relation of $\gamma$ matrices. Correspondingly,
the Lagrangian is ${\cal L}=\bar{\psi}(i\kappa\slashed D-m)\psi-\frac{1}{4}\eta^2(F^\a_{\mu\nu})^2$ with $D_\mu=\p_\mu+iec_2\gamma^\nu(\gamma\p^{\b-\a})_\nu A_\mu(x)$. Since physical electron charge $e=c_2e_0$, once there is non-local interaction $\a\neq\b$, it will alter the mass dimension of $e$ by $m^{\a-\b}$. We can observe the mass dimension of electron charge to examine if there exist non-local interaction from Le'vy paths.

From above discussion, Dirac spinor field in vacuum with Le'vy paths is nonphysical. While Le'vy path electromagnetic field in vacuum can have gauge interaction with electron and thus may have observable effects, such as non-locality in light-matter interaction and modification on electron charge.

\section{conclusion}\label{summary}
We introduce Le'vy path integral approach to quantum field in several scenes including scalar field with three dimensional box boundary conditions. We obtain its density of state and find it describes systems with memories on field strength, which is dual to lower dimensional systems with no memories. While for  $\a<\frac{1}{3}$ case, the density of states is dual to a system with less than one dimension and has very high degeneracy near frequency $\omega=0$. As an example, we calculate corresponding spectrum of black body radiation. It can dual to vary dimensional black body. In zero temperature fractional quantum fields, we find that its property in $\a\rightarrow1$ can be used as a technique to regular divergence in one loop Feynman diagram as Dimensional Regularization. Furthermore, we study the existence of Le'vy paths in zero temperature fractional Dirac spinor field and electromagnetic field from perspective of gauge symmetry and stability. We find Le'vy paths are forbidden in Dirac spinor field by gauge symmetry and will led to unstable electrons. But we find a nonlocal phase factor and corresponding nonlocal interaction which permit Le'vy paths in fractional electromagnetic field, thus may have observable effects, such as nonlocality in light-matter interaction and modification on electron charge.

\section{Appendix}
\label{ioop}
For integral 
\begin{equation}
\begin{split}
	\label{eq21}I=(-i)^{-\a}\eta^{-1}\int_{-\infty}^{+\infty} d^4k\frac{2i\slashed{k}^\a-4m_1}{k^{2\a}+m_1^2+i\epsilon}\frac{1}{(k-p)^{2\a}+i\epsilon},\qquad m_1=m/(\eta (-i)^\a)\,,
\end{split}
\end{equation}
use 
\begin{equation}
\begin{split}
	(k-p)^{2\alpha}=k^{2\alpha}-2\alpha pk^{2\alpha-1}+b,\quad b=\begin{cases}0\quad&\alpha\neq1\\p^{2}\quad&\alpha=1,
		\end{cases}
\end{split}
\end{equation}
and 
\begin{equation}
\begin{split}
	\frac{1}{k^2+m^2+i\epsilon}\frac{1}{(k-p)^2+i\epsilon}=\int_{0}^{1}dx\frac{1}{[l^2+x(1-x)p^2+(1-x)m^2+i\epsilon]^2}\,,
\end{split}
\end{equation}
we have
\begin{equation}
\begin{split}
	\frac{1}{k^{2\alpha}+m_1^2+i\epsilon}\frac{1}{(k-p)^{2\alpha}+i\epsilon}=\int_{0}^{1}dx\frac{1}{[(k-xp)^{2\alpha}+\triangle+i\epsilon]^{2}},
\end{split}
\end{equation}
where $\triangle=x(1-x)b+(1-x)m_1^2$. Substitute it into Eq.~(\ref{eq21}) and use transition $l=k-xp$. For $(l+xp)^\a$ in numerator, we make use of
\begin{equation}
\begin{split}
	(l+xp)^\alpha&=\bar{\partial}^{1-\alpha}(l+xp)^1\\&=l^{\alpha}+b,\ \ \mbox{where} \ \ c=\begin{cases}0\quad&\alpha\neq1\\xp\quad&\alpha=1\,.
	\end{cases}
 \label{b1}
\end{split}
\end{equation}
We just focus on the integral of the momentum
\begin{equation}
\begin{split}
I_1&=\int d^4k\frac{2i\cancel{(l^{\a}+b)}-4m_1}{[l^{2\alpha}+\triangle+i\epsilon]^{2}}\\&=2\int d^4k\frac{i\slashed{z}^{-1}(l^{\a}+b)_3-2m_1}{[l^{2\alpha}+\triangle+i\epsilon]^{2}}\,,
\end{split}
\end{equation}
where $\slashed{z}=\gamma^{\mu}n_{\mu}$, $n_{\mu}=z$ is a unit vector along the $z$ axis. We split this integral into two parts: 
\begin{equation}
\begin{split}
	I_1=2(i\slashed{z}^{-1}I_3-2m_1I_2).
\end{split}
\end{equation}

\subsubsection{Compute $I_2$ and $I_3$}
\label{I}
{\bf Integral $I_2$}

Let $l_{0}=e^{\frac{i\pi}{2\alpha}}l_E, l_0^{2\a}=-l_E^{2\a}$, $l_{i}=l_E, (i=1,2,3)$, and make quasi Wick rotation.
\begin{equation}
\begin{split}
	I_2&=e^{\frac{i\pi}{2\alpha}}\int d^{4}l_E\frac{1}{(l_{E}^{2\alpha}+\triangle+i\epsilon)^{2}}\\&=\frac{e^{\frac{i\pi}{2\alpha}}}{2\alpha}\int d\Omega\int dl_{r}^{2\alpha}\frac{l_{r}^{4-2\alpha}}{(l_{r}^{2\alpha}\xi+\triangle+i\epsilon)^{2}}\\&=\frac{e^{\frac{i\pi}{2\alpha}}}{2\alpha}\int d\Omega(\xi^{-\frac{2}{\alpha}})\int_{0}^{\infty}dr\frac{r^{\frac{2-\alpha}{\alpha}}}{(r+\triangle+i\epsilon)^{2}}\\&=\frac{e^{\frac{i\pi}{2\alpha}}}{2\alpha}\int d\Omega(\xi^{-\frac{2}{\alpha}})\triangle^{\frac{2-2\alpha}{\alpha}}\frac{\Gamma(\frac{2\alpha-2}{\alpha})\Gamma(\frac{2}{\alpha})}{\Gamma(2)},
\end{split}
\end{equation}
where $l_{r}$ is radial coordinate and $r=l_{r}^{2\alpha}\xi$, 
$$\xi=\cos^{2\alpha}\theta+\sin^{2\alpha}\theta[\cos^{2\alpha}\phi+\sin^{2\alpha}\phi(\cos^{2\alpha}\omega+\sin^{2\alpha}\omega)].$$ 
We haven't  integrate $d\Omega=sin^2\theta~sin\phi~d\theta~d\phi ~d\omega$ successfully yet. Anyway, when $\alpha=1$, the angle integral is simple to be $2\pi^2$.

{\bf Integral $I_3$}

Use the same process we can get the result.
\begin{equation}
\begin{split}
	I_3=\frac{1}{2\alpha}\int d\Omega(\cos^{\alpha}\theta \xi^{-\frac{4+\alpha}{2\alpha}})\triangle^{\frac{4-3\alpha}{2\alpha}}\frac{\Gamma(\frac{3}{2}-\frac{2}{\alpha})\Gamma(\frac{1}{2}+\frac{2}{\alpha})}{\Gamma(2)}+b_3I_2,
\end{split}
\end{equation}
where $b$ is denoted in Eq.~(\ref{b1}).
When $\alpha=1$, $\cos^{\alpha}\theta~(\theta\in[0,\pi])$ will make the angle integral vanish, so $I_3=b I_2$, and the whole loop integral goes back to the ordinary case. 
\begin{acknowledgements}
The authors thank Lu Yang and  Jun Chang for helpful discussions. This work is supported by the National Natural Science Foundation of China under Grant No.~11847168, the Fundamental Research Funds for the Central Universities of China under Grant No.~GK202003018, and the Natural Science Foundation of Shannxi Province, China (2019JM-431, 2019JQ-739). 
\end{acknowledgements}

\bibliographystyle{apsrev4-1}
%\bibliography{gravreferences}

\end{document}